\documentclass[prd,twocolumn,floatfix,amsmath,nofootinbib,amssymb,floatfix]{revtex4}
\usepackage{graphicx,color,dcolumn,booktabs,bm}
\usepackage{longtable,lscape}
\usepackage{txfonts}
\usepackage{overpic}
\usepackage{amssymb}
\usepackage{indentfirst}
\usepackage{feynmf}   
\usepackage{slashed}  
\usepackage{cases}
\usepackage{color}
\usepackage{multirow}
\usepackage{epstopdf}
\usepackage{enumerate}
\usepackage{graphicx,color,dcolumn,booktabs,bm}
\usepackage[colorlinks, citecolor=blue,anchorcolor=red,menucolor=red, linkcolor=red,filecolor=red,urlcolor=blue,frenchlinks=red]{hyperref}

\graphicspath{{Figures/}} %

\begin{document}

\title{Toward $e^+e^-\to \pi^+\pi^-$ annihilation inspired by higher $\rho$ mesonic states  around 2.2 GeV}
\author{Li-Ming Wang$^{1,2}$}\email{lmwang15@lzu.edu.cn}
\author{Jun-Zhang Wang$^{1,2}$}\email{wangjzh2012@lzu.edu.cn}
\author{Xiang Liu$^{1,2,3}$\footnote{Corresponding author}}\email{xiangliu@lzu.edu.cn}
\affiliation{$^1$School of Physical Science and Technology, Lanzhou University, Lanzhou 730000, China\\
$^2$Research Center for Hadron and CSR Physics, Lanzhou University $\&$ Institute of Modern Physics of CAS, Lanzhou 730000, China\\$^3$Joint Research Center for Physics, Lanzhou University and Qinghai Normal University, Xining 810000, China}

\begin{abstract}

Very recently, the {\it BABAR} Collaboration indicated that there exist an explicit enhancement structure near 2.2 GeV when focusing on the $e^+e^-\to\pi^+\pi^-$ process again, which inspires our interest in studying the production of higher $\rho$ mesonic states. Since the branching ratio of $\pi^+\pi^-$ channel of $D-$wave $\rho$ states are much smaller than $S-$wave states, we choose $\rho(1900)$ and $\rho(2150)$ as the intermediate states in $e^+e^-\to\pi^+\pi^-$, where $\rho(1900)$ and $\rho(2150)$ are treated as $\rho(3S)$ and $\rho(4S)$ states, respectively. Our result indicates that the $BABAR$'s data of $e^+e^-\to\pi^+\pi^-$ around 2 GeV can be depicted well, which shows that this enhancement structure near 2.2 GeV existing in $e^+e^-\to\pi^+\pi^-$ can be due to the contribution from two $\rho$ mesons, $\rho(1900)$ and $\rho(2150)$. Additionally, this conclusion can be enforced by the consistence of the extracted values of $\Gamma_{e^+e^-}\mathcal{B}(\pi^+\pi^-)$ of  $\rho(1900)$ and $\rho(2150)$ in the whole fitting processes and the corresponding theoretical calculations. The present study of $e^+e^-\to\pi^+\pi^-$ data may provide valuable information to establish the $\rho$ meson family.
\end{abstract}
\date{\today}
\maketitle

\section{\label{sec1}Introduction}

$e^+e^-$ annihilation process can be as an ideal platform to study vector particles. A typical example is the observation of $J/\psi$
charmonium \cite{Augustin:1974xw}. By adopting initial state radiation method which plays crucial role to the observation of charmoniumlike state $Y(4260)$ from the $e^+e^-\to J/\psi\pi^+\pi^-$ process \cite{Aubert:2005rm}, the {\it BABAR} Collaboration measured the $e^+e^-\to \pi^+\pi^-$ process
\cite{Lees:2012cj} by the collected 232 fb$^{-1}$ experimental data, by which the cross section information of $e^+e^-\to \pi^+\pi^-$ from the $\pi^+\pi^-$ threshold to center-of-mass energy of 3 GeV was obtained.  Very recently, the {\it BABAR} Collaboration focused on $e^+e^-\to \pi^+\pi^-$ again, and indicated that there exists an explicit enhancement structure near 2.2 GeV in the $\pi^+\pi^-$ invariant mass spectrum \cite{BABAR:2019oes}. This phenomenon stimulates our interest in studying $e^+e^-\to \pi^+\pi^-$ since it has a close relation to establish light vector mesons around 2 GeV, which is one part of whole study of light hadron spectroscopy.

In fact, the $\pi^+\pi^-$ final state determines that $e^+e^-\to \pi^+\pi^-$ is a clean process to explore light vector $\rho$ mesons with positive $G$ parity. Generally, light vector mesons can be grouped into isovector $\rho$ meson family, isoscalar $\omega$ and $\phi$ meson families.
If checking the mass spectrum of light vector meson \cite{Wang:2012wa,He:2013ttg,Wang:2019jch}, we may find that some higher $\rho$, $\omega$, and $\phi$ states accumulate around 2 GeV mass range, which may result in the  difficulty of distinguishing them when analyzing some annihilation processes of $e^+e^-$ into light mesons.
It is obvious that the pollution from $\omega$ and $\phi$ mesons can be avoided for $e^+e^-\to \pi^+\pi^-$, which is the reason why we are dedicated to the study of $e^+e^-\to \pi^+\pi^-$  by combing with higher $\rho$ mesons around 2 GeV.

In Ref. \cite{He:2013ttg}, Lanzhou group once performed the mass spectrum analysis and calculated these two-body Okuba-Zweig-Iizuka (OZI)  allowed decays of $\rho$ meson family.  By combining with these reported $\rho$-like states collected in Particle Data Group (PDG) \cite{Tanabashi:2018oca},
the possible assignment of these $\rho$-like states into the $\rho$ meson family was suggested \cite{He:2013ttg}, which is crucial step of constructing $\rho$
meson family. However, it is not the end of whole story.

After releasing the detailed data of cross section of $e^+e^-\to \pi^+\pi^-$ \cite{Lees:2012cj,BABAR:2019oes} by {\it BABAR}, we may continue to
carry out the study of the production of higher $\rho$ mesonic states around 2 GeV via $e^+e^-\to \pi^+\pi^-$. Based on the results given in Ref. \cite{He:2013ttg}, we may select suitable higher $\rho$ mesons around 2 GeV as the intermediate states in $e^+e^-\to \pi^+\pi^-$.
And then, by fitting the cross section of $e^+e^-\to \pi^+\pi^-$ around 2.2 GeV under our theoretical approach, we may get the information of the contribution of these $\rho$ mesonic states to the cross section of $e^+e^-\to \pi^+\pi^-$, which is valuable to
further test the suggestive assignment of the $\rho$ meson family \cite{He:2013ttg}.
We hope that our effort presented in this work can improve our understanding of constructing the $\rho$ meson family, especially for these higher $\rho$ mesonic states around 2 GeV.

This paper is organized as follows. After Introduction, we will give a concise review of $\rho$ mesons with mass around 2 GeV, and introduce
the possible assignment to them (see Sec. \ref{sec2}). In Sec. \ref{sec3}, we will present our theoretical framework of calculating $e^+e^-\to \pi^+\pi^-$ with these discussed $\rho$ mesons as intermediate states. When fitting the {\it BABAR} data, we finally extract the magnitude of different $\rho$ meson contributions to the $e^+e^-\to \pi^+\pi^-$ cross section around 2.2 GeV. And then, we give the numerical result in  Sec. \ref{sec4}. This paper ends with the summary.

\section{The situation of $\rho$ mesons around 2 GeV}\label{sec2}

There are many $\rho$-like states reported by experiments. Among these states, $\rho(770)$ is a well established ground state with very broad width. As shown in PDG \cite{Tanabashi:2018oca}, $\rho(1450)$ can be assigned as the first radial excited state of $\rho(770)$. By the analysis of the mass spectrum \cite{Godfrey:1985xj} and the study of total decay width \cite{He:2013ttg} and the branching ratio of the $\rho(1700)\to 2\pi, 4\pi$ \cite{Abele:2001pv} and $e^+e^-\to\omega\pi^0$ process \cite{Kittimanapun:2008wg}, we may find that $\rho(1700)$ as a candidate of the $\rho(1^3D_1)$ meson state is suitable. It is the research status of some low-lying $\rho$-like states.

$\rho(1900)$ was firstly observed by the DM2 Collaboration, which corresponds to a dip around 1.9 GeV with analyzing the process $e^+e^-\to 6\pi$ \cite{Castro:1988hp}. After that, there were many experiments relevant to $\rho(1900)$ which include the FENICE Collaboration \cite{Antonelli:1996xn}, the E687 Collaboration \cite{Frabetti:2001ah,Frabetti:2003pw}, the $BABAR$ Collaboration \cite{Aubert:2006jq,Aubert:2007ym}, and the CMD3
Collaboration \cite{Solodov:2011dn}. In addition, the analysis of experimental data on isovector $P-$wave of pion-pion scattering also testifies in favor of the existence of $\rho(1900)$ \cite{Surovtsev:2008zza}. It is worth nothing that $\rho(1900)$ was identified from $6\pi$ peak exactly at the $p\bar{p}$ threshold \cite{Solodov:2011dn}. Thus, Bugg suggested that this state is likely to be a $\rho(^3S_1)$ captured by the very strong $p\bar{p}$ $S-$wave but could be a nonresonant cusp effect \cite{Bugg:2012yt}.

In 2013, Lanzhou group \cite{He:2013ttg} indicated that $\rho(1900)$ can be regarded as $3^3S_1$ state since the obtained total width overlaps with the $BABAR$'s dada \cite{Barnes:1996ff}. Here, the main decay channels of $\rho(1900)$ are $\pi\pi$, $\pi a_1(1260)$, $\pi h_1(1170)$, $\pi\pi(1300)$, and $\pi\omega(1420)$ \cite{He:2013ttg}. Therefor, $\rho(1900)$ with a large branching ratio of $4\pi$ can be understood.

 Clegg and Donnachi jointly analyzed the data on $6\pi$ states produced in the $e^+e^-$ annihilation and diffractive photoproduction, and indicated that there exists a resonance with peak near 2.1 GeV \cite{Bisello:1981sh,Atkinson:1985yx,Clegg:1989mp}, which corresponds to $\rho(2150)$. Later, $\rho(2150)$ was assigned as the third radial excitation of $\rho(770)$ by fitting the pion form factor \cite{Biagini:1990ze}. In addition, other experiments like GAMS \cite{Alde:1992wv,Alde:1994jm}, Crystal Barrel \cite{Anisovich:2002su,Anisovich:2000ut,Anisovich:2001vt,Anisovich:1999xm} and $BABAR$ \cite{Aubert:2007ef} confirmed the observation of $\rho(2150)$ in different processes.

According to the analysis of mass spectrum \cite{Bugg:2012yt,Anisovich:2000kxa,Masjuan:2012gc,Masjuan:2013xta}, $\rho(2150)$ can be a good candidate of $\rho(4^3S_1)$ meson state. The study of OZI-allowed two-body strong decay behaviors of $\rho(2150)$ supports this assignment \cite{He:2013ttg}, since the SPEC's data  \cite{Anisovich:2002su} can be reproduced  \cite{He:2013ttg}. Here, the dominant channels of $\rho(2150)$ are $\pi\pi$, $\pi a_1(1260)$, $\pi\omega$ and $\pi h_1(1170)$ \cite{He:2013ttg}, which can explain why $\rho(2150)$ was observed in $\pi^+\pi^-$, $\omega\pi^0$, $\eta'\pi\pi$, $f_1(1285)\pi\pi$, and $\omega\pi\eta$ experimentally.

\begin{table}[htbp]
	\caption{The suggested assignment to these observed $\rho$-like states and the information of their main decay channels from Ref. \cite{He:2013ttg}.\label{rho}}
	\begin{center}
		\renewcommand{\arraystretch}{1.7}
		\tabcolsep=1.2pt
\setlength{\tabcolsep}{2mm}
{
		\begin{tabular}{ccc}
			\toprule[1pt]\toprule[1pt]
State &Assignment        &Main decay channels \\
\midrule[1pt]
$\rho(770)$   & $\rho(1S)$ & $\pi\pi$ \\
$\rho(1450)$  & $\rho(2S)$ & $\pi\pi$, $\pi a_1(1260)$, $\pi\omega$, $\pi h_1(1170)$ \\
$\rho(1900)$  & $\rho(3S)$ & $\pi\pi$, $\pi a_1(1260)$, $\pi h_1(1170)$,\\
&&$\pi\pi(1300)$, $\pi\omega(1420)$ \\
$\rho(2150)$  & $\rho(4S)$ & $\pi\pi$, $\pi a_1(1260)$, $\pi\omega$, $\pi h_1(1170)$ \\
$\rho(1700)$  & $\rho(1D)$ & $\pi a_1(1260)$, $\pi h_1(1170)$ \\
$\rho(2000)$  & $\rho(2D)$ & $\pi\pi(1300)$, $\rho\rho$, $\pi\pi_2(1670)$, $\pi a_1(1260)$ \\
$\rho(2270)$  & $\rho(3D)$ & $\pi\pi(1300)$, $\pi\pi(1800)$ \\
       \bottomrule[1pt]\bottomrule[1pt]
		\end{tabular}
}
	\end{center}
\end{table}

In PDG \cite{Tanabashi:2018oca}, there are two $\rho$-like states are listed as further state, which are $\rho(2000)$ and $\rho(2270)$. $\rho(2000)$ was observed in the $p\bar{p}\to\pi\pi$ reaction with the mass around 1988 MeV \cite{Hasan:1994he}. Later, a combined fit was presented to the data of $p\bar{p}\to \omega\eta\pi^0$ and $\omega\pi$, by which the existence of $\rho(2000)$ was confirmed \cite{Anisovich:2000ut}. The analysis of the Regge trajectory shows $\rho(2000)$ as the first radial excitation of $\rho(1700)$ \cite{Anisovich:2002su,He:2013ttg}. The dominant channels of $\rho(2000)$ include $\pi\pi(1300)$, $\rho\rho$, $\pi\pi_2(1670)$ and $\pi a_1(1260)$ indicated in Ref. \cite{He:2013ttg}.

In the reaction $\gamma p\to\omega\pi^+\pi^-\pi^0$, a spin-parity analysis shows the existence of a resonance with $J^P=1^-$ in the $\omega\rho^{\pm}\pi^{\mp}$ final state, which has mass around 2.28$\pm$0.05 GeV \cite{Atkinson:1985yx}. And then, the Crytal Barrel experiment fitted the $\omega\eta\pi$ data from the $p\bar{p}$ annihilation, where $\rho(2270)$ was confirmed \cite{Anisovich:2000ut}. The Regge trajectory analysis gives that $\rho(2270)$ is a good candidate of the second radial excitation of $\rho(1700)$ \cite{Anisovich:2002su,He:2013ttg}. The decay information of $\rho(2270)$ was also provided in Ref. \cite{He:2013ttg}.

In Table \ref{rho}, we collect the information of these reported $\rho$-like states, and their assignments and dominant decay channels.

\section{Depicting the cross section of $e^+e^-\to \pi^+\pi^-$ around 2 GeV}\label{sec3}

\begin{figure}[hbtp]
\centering
\includegraphics[width=0.49\textwidth,scale=0.36]{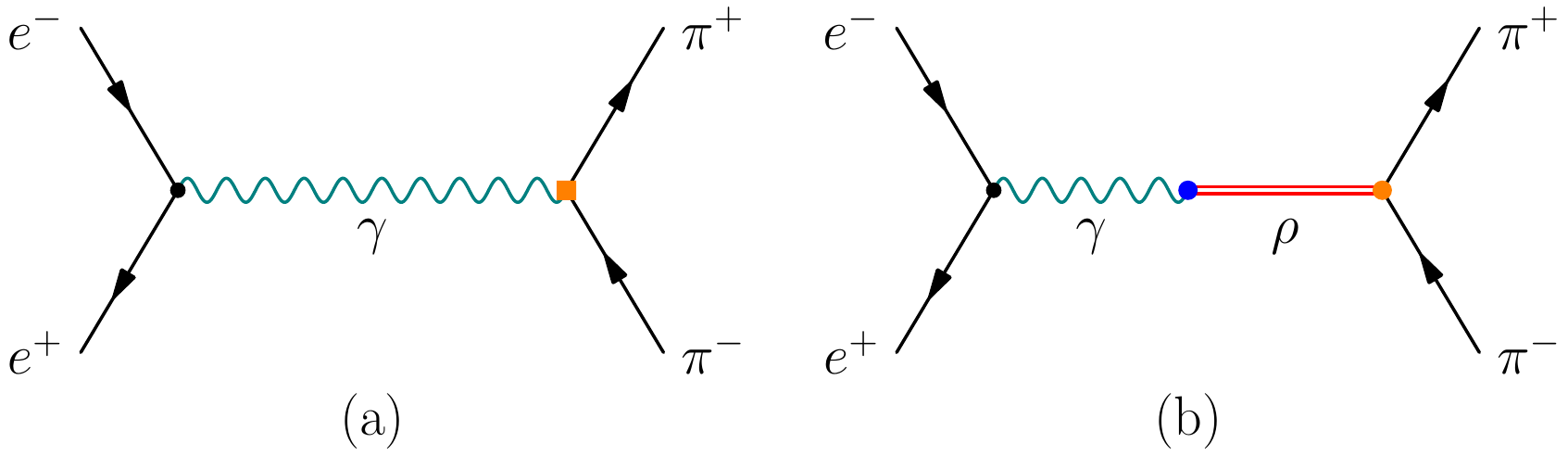}
\caption{(Color online.) The diagrams  for  depicting the $e^+e^-\to\pi^+\pi^-$ process. Here, (a) is direct annihilation process while (b) corresponds to the intermediate $\rho$ state contribution.}\label{lag}
\end{figure}

In this section, we focus on the $e^+e^-\to \pi^+\pi^-$ process at center-of-mass energy around 2 GeV. Due to the constraint of conservation of $G$ parity, $e^+e^-\to \pi^+\pi^-$ can be applied to study
$\rho$-like states. As shown in Fig. \ref{lag}, there exist two mechanisms working together for $e^+e^-\to \pi^+\pi^-$. The first one is $e^+e^-$ direct annihilation into $\pi^+\pi^-$, where the virtual photon couples with the final state $\pi^+\pi^-$, which provides the background contribution. The second one is that $e^+e^-\to \pi^+\pi^-$ occurs via intermediate $\rho$ states.

When trying to reproduce the line shape of the cross section of $e^+e^-\to\pi^+\pi^-$ process around 2 GeV reported by the $BABAR$ Collaboration recently \cite{BABAR:2019oes}, we need to choose suitable intermediate $\rho$ meson states. The collected information in Table \ref{rho} shows that $\rho(1900)$, $\rho(2000)$, $\rho(2150)$ and $\rho(2270)$ should be considered in our calculation.

In this work, we adopt effective Lagrangian approach to calculate these discussed processes shown in Fig. \ref{lag}.
The effective Lagrangian involved in the concrete work include  \cite{Chen:2020xho}
\begin{eqnarray}\label{Lag}
\begin{split}
\mathcal{L}_{\pi\pi\gamma}=&ieA^{\mu}(\pi\partial_{\mu}\pi-\partial_{\mu}\pi\pi),\\
\mathcal{L}_{\rho\pi\pi}=&ig_{\rho\pi\pi}\rho_i^{\mu}(\pi\partial_{\mu}\pi-\partial_{\mu}\pi\pi),\\
\mathcal{L}_{\gamma\rho}=&-e\frac{m_{\rho}^2}{f_{\rho}}\rho^{\mu}A_{\mu}.
\end{split}
\end{eqnarray}
It is worth noting that the above Lagrangian densities are not unique. For example, for $\mathcal{L}_{\rho\pi\pi}$, the Lorentz structure involving the derivative of the field-strength tensor of the $\rho$ meson field, i.e., $\partial_{\mu}F_{\mu\nu}(\pi\partial_{\nu}\pi)$ with $F_{\mu\nu}=\partial_{\mu}\rho_{\nu}-\partial_{\nu}\rho_{\mu}$, is also allowed. However, the current experimental data from {\it BABAR} does not support us to consider more Lagrangian densities in a realistic calculation. Of course, we hope that these contributions from other Lagrangian couplings can be included in accurate experimental measurements of $e^+e^-\to \pi^+ \pi^-$ in the future.

The amplitudes corresponding to the diagrams in Fig. \ref{lag} can be written as
\begin{eqnarray}\label{Mag}
\begin{split}
\mathcal{M}_{\mathrm{Dir}}=&\left[\bar{v}(p_2,m_e)(ie\gamma^{\mu})u(p_2,m_e)
\right]\frac{-g_{\mu\nu}}{q^2}
\left[ie(p_4^{\nu}-p_3^{\nu})F_{\pi}(q^2)
\right],\\
\mathcal{M}_{\rho_i}=&\left[\bar{v}(p_2,m_e)(ie\gamma_{\mu})u(p_2,m_e)
\right]\frac{-g^{\mu\xi}}{q^2}\left(-e\frac{m_{\rho_i}^2}{f_{\rho_i}^2}\right)\\
&\times \frac{-g_{\xi\nu}+q_{\xi}q_{\nu}/m_{\rho_i}^2}{q^2-m_{\rho_i}^2+im_{\rho_i}\Gamma_{\rho_i}}
\left[ig_{\rho_i\pi\pi}(p_4^{\nu}-p_3^{\nu})\right].
\end{split}
\end{eqnarray}
Here, $F_{\pi}$ is the time-like form factor of charged pion and $q=p_1+p_2$.\;$\rho_i$ denote intermediate $\rho$ meson states.\;$m_{\rho_{i}}$ and $\Gamma_{\rho_{i}}$ are resonance parameter, which can be fixed by the corresponding experimental data  \cite{Tanabashi:2018oca}. The total amplitude of $e^+e^-\to\pi^+\pi^-$ is superposition of different contribution
\begin{eqnarray}\label{MT}
\begin{split}
\mathcal{M}_{\mathrm{Total}}=\mathcal{M}_{\mathrm{Dir}}+\sum_i e^{i\theta_i}\mathcal{M}_{\rho_i},
\end{split}
\end{eqnarray}
where $\theta_i$ denotes the phase angle between the amplitudes from direct annihilation and  the intermediate $\rho$ state contribution. $g_{\rho_i}=g_{\rho_i\pi\pi}m_{\rho_i}^2/f_{\rho_i}$ and $f_{\rho_i}$ represents decay constant of some $\rho$ meson. With the above amplitude, the differential cross section of $e^+e^-\to\pi^+\pi^-$ can be calculated directly, i.e.,
\begin{eqnarray}\label{cs}
\begin{split}
\frac{\mathrm{d}\sigma}{\mathrm{d}t}=\frac{1}{64\pi s}\frac{1}{|p_{\mathrm{1cm}}|^2}\overline{|\mathcal{M}_{\mathrm{Total}}|^2}.
\end{split}
\end{eqnarray}

When fitting the cross section for $e^+e^-\to\pi^+\pi^-$, we can treat $\theta_i$ and $g_{\rho_i}$ as free parameters.
Additionally, the form factor of charged pion is not determined since the form factor in the time-like range is a complex function. In general, the form factor changes slowly when $q^2$ is far away from threshold. For simplicity, we assume the form factor is a constant in the center-of-mass energy considered here. Thus, we argue that the form factor of charged pion can be absorbed into phase angle\footnote{Assuming that the pion form factor is the product of  a constant term and an exponential term, and complex character of form factor can be reflected in the constant term, i.e.,
\begin{eqnarray}\label{MT1}
\begin{split}
F_{\pi}(s)=ae^{-i\phi}e^{-b\sqrt{s}},
\end{split}
\end{eqnarray}
where $e^{-i\phi}$ is an any complex number and its modulus can be absorbed into free parameter $a$. According to the definitions in Eqs. (\ref{Mag})-(\ref{MT}), then
the total amplitude can be written as
\begin{eqnarray}\label{MT2}
&&\mathcal{M}_{\mathrm{Total}}=e^{-i\phi}\mathcal{M}_{\mathrm{Dir}}+\sum_i e^{i\phi_i}\mathcal{M}_{\rho_i} \nonumber  \\
&&\Longrightarrow
e^{-i\phi}\mathcal{M}_{\mathrm{Total}}^{\prime}=e^{-i\phi}(\mathcal{M}_{\mathrm{Dir}}+\sum_i e^{i(\phi_i+\phi)}\mathcal{M}_{\rho_i}).
\end{eqnarray}
Because $|e^{-i\phi}\mathcal{M}_{\mathrm{Total}}^{\prime}|^2=|\mathcal{M}_{\mathrm{Total}}|^2=|\mathcal{M}_{\mathrm{Total}}^{\prime}|^2$, so
the above scattering amplitude is equivalent to Eq. (\ref{MT}).
That is to say, the complex part of form factor can be absorbed into phase angle between the amplitudes from direct annihilation and  the intermediate $\rho$ state contribution. Thus, the form factor can be taken as a real form.}. In the present work, the form factor is taken as
$F_K(s)=ae^{-b\sqrt{s}}$,
where $a$ and $b$ are free parameters.

By the effective Lagrangian listed in Eq. (\ref{Lag}), the dilepton and $\pi^+\pi^-$ decay widths of these intermediate $\rho$ meson state can be expressed as
\begin{eqnarray}\label{g}
\begin{split}
\Gamma_{e^+e^-}=&\frac{e^4m_{\rho_i}}{12\pi f_{\rho_i}^2},\\
\Gamma_{\pi^+\pi^-}=&\frac{g_{\rho_i\pi\pi}^2(m_{\rho_i}^2-4m_{\pi}^2)^{3/2}}{48\pi m_{\rho_i}^2},
\end{split}
\end{eqnarray}
respectively. Thus, we may further define  the production of dilepton decay width and the branching ratio of $\pi^+\pi^-$ mode of the discussed $\rho$ meson
\begin{eqnarray}\label{eepipi}
\Gamma_{e^+e^-}\mathcal{B}({\pi^+\pi^-})=\frac{e^4g_{\rho_i}^2(m_{\rho_i}^2-4m_{\pi}^2)^{3/2}}{576\pi^2 m_{\rho_i}^5\Gamma_{\rho_i}},
\end{eqnarray}
where $m_{\rho_i}$ and $m_{\pi}$ are the masses of intermediate $\rho$ states and final state pion, respectively.

\begin{table*}[htbp]
	\caption{The information of four intermediate $\rho$ state involved in the $e^+e^-\to \pi^+\pi^-$ process around 2.2 GeV. The second and the third columns are resonance parameter. $R$ is the parameter in SHO wave function (see Eq. (\ref{L2})) which was given in Ref. \cite{He:2013ttg}. $\mathcal{B}(\pi^+\pi^-)$ is branching ratio of $\pi^+\pi^-$ mode calculated via the QPC model \cite{He:2013ttg}. $\Gamma_{e^+e^-}$ is dilepton decay width calculated in this work by Eq. (\ref{eepipit}). By the numerical results listed in the fifth and the sixth columns,  the results of $\Gamma_{e^+e^-}\mathcal{B}(\pi^+\pi^-)$  can be obtained.}\label{chidu}
	\begin{center}
		\renewcommand{\arraystretch}{2.0}
		\tabcolsep=1.2pt
\setlength{\tabcolsep}{2.5mm}
{
		\begin{tabular}{c|cccccc}
\toprule[1pt]\toprule[1pt]
state & $M_{\mathrm{exp}} $ (MeV) & $\Gamma_{\mathrm{exp}}$ (MeV)  &  $R$ (GeV$^{-1}$)  \cite{He:2013ttg} & $\mathcal{B}(\pi^+\pi^-)$ \cite{He:2013ttg} & $\Gamma_{e^+e^-}$ (keV) & $\Gamma_{e^+e^-}\mathcal{B}(\pi^+\pi^-)$ (keV)\\

\midrule[1pt]
$\rho(1900)$  & $1909\pm17\pm25$ \cite{Aubert:2006jq} & $160\pm20$ \cite{Aubert:2006jq}  & $3.85\sim 4.28$ & $0.1450\sim0.3509$ & $0.1958\sim 0.1578$  & $0.0284\sim 0.0554$ \\

$\rho(2150)$  & $2150\pm17$ \cite{He:2013ttg}  & $230\pm50$ \cite{Anisovich:2002su} & $4.74\sim4.98$ & $0.3889\sim0.3396$ & $0.0888\sim0.0806$  & $0.0345\sim0.0274$ \\

$\rho(2000)$   & $2000\pm30$ \cite{Bugg:2003kj}  & $260\pm45$ \cite{Bugg:2003kj}  & $4.34\sim4.80$ & $0.0740\sim0.0573$ & $0.0204\sim0.0160$ & $0.0015\sim0.0009$ \\

$\rho(2270)$ & $2265\pm40$ \cite{Anisovich:2002su}  & $325\pm80$ \cite{Anisovich:2002su} &$4.40\sim4.80$   & $0.0510\sim0.0315$ & $0.0163\sim0.0129$& $0.0008\sim0.0004$ \\
\bottomrule[1pt]\bottomrule[1pt]
       \end{tabular}}

	\end{center}
\end{table*}

\begin{table}[htbp]
	\caption{The parameters obtained by fitting the cross section of $e^+e^-\to\pi^+\pi^-$ measured by {\it BABAR} \cite{BABAR:2019oes}. }\label{para}
	\begin{center}
		\renewcommand{\arraystretch}{1.5}
		\tabcolsep=1.2pt
\setlength{\tabcolsep}{2.0mm}
{
		\begin{tabular}{c|cc}
\toprule[1pt]\toprule[1pt]
          Parameters                                              & Solution A    & Solution B  \\  
\midrule[1pt]
$a$                                                     & $0.22\pm0.16$ & $0.54\pm0.02$ \\
$b$ (GeV$^{-1}$)                                        & $1.06\pm0.36$ & $1.37\pm0.01$ \\
$\theta_1$ (rad)                                        & $1.55\pm0.49$ & $1.45\pm0.35$ \\
$\theta_2$ (rad)                                        & $4.98\pm0.10$ & $5.17\pm0.18$ \\
$\Gamma_{e^+e^-}\mathcal{B}(\pi^+\pi^-)_{\rho(1900)}$ (eV)&$15.06\pm6.10$&$3.80\pm0.89$\\
$\Gamma_{e^+e^-}\mathcal{B}(\pi^+\pi^-)_{\rho(2150)}$ (eV)&$44.53\pm18.33$&$2.74\pm0.79$\\
$\chi^2/\mathrm{d.o.f}$                                 & 1.10          & 1.06        \\
\bottomrule[1pt]\bottomrule[1pt]
       \end{tabular}}

	\end{center}
\end{table}

\section{NUMERICAL RESULTS}\label{sec4}

In the following, we will fit the cross section for $e^+e^-\to\pi^+\pi^-$ measured by the {\it BABAR} Collaboration \cite{BABAR:2019oes}, where an event accumulation near 2.2 GeV exists in the $\pi^+\pi^-$ invariant mass spectrum. As seen in Table \ref{chidu}, $\pi^+\pi^-$ is the most dominant decay channel for the $S$-wave $\rho$ mesons but not important to the $D$-wave $\rho$ mesons, which is from the theoretical calculation in Ref. \cite{He:2013ttg}.  Especially, we find that the theoretical result of $\Gamma_{e^+e^-}\mathcal{B}(\pi^+\pi^-)$
for $S$-wave $\rho$ mesons ($\rho(1900)$ and $\rho(2150)$) is one order of magnitude larger than that for $D$-wave $\rho$ mesons ($\rho(2000)$ and $\rho(2270)$). Thus,
in order to reduce the number of fitting parameters,  $\rho(2000)$ and $\rho(2270)$ that are treated as $D$-wave meson states \cite{He:2013ttg} will not be considered in the following study. In our realistic analysis, we only choose $\rho(1900)$ and $\rho(2150)$ as intermediate resonances, which are assigned as $\rho(3S)$ and $\rho(4S)$ states, respectively.
When fitting the experimental data under our theoretical framework, there are six free parameters, $a$, $b$, $\theta_1$, $\theta_2$, $\Gamma_{e^+e^-}\mathcal{B}(\pi^+\pi^-)_{\rho(1900)}$ and $\Gamma_{e^+e^-}\mathcal{B}(\pi^+\pi^-)_{\rho(2150)}$.
Here, the subscripts in $\Gamma_{e^+e^-}\mathcal{B}(\pi^+\pi^-)_{\rho(1900)}$ and $\Gamma_{e^+e^-}\mathcal{B}(\pi^+\pi^-)_{\rho(2150)}$ is applied to distinguish the $\rho(1900)$ and $\rho(2150)$ contributions.
Additionally, [$m_{\rho(1900)}$, $\Gamma_{\rho(1900)}$] and [$m_{\rho(2150)}$, $\Gamma_{\rho(2150)}$] as the resonance parameters of $\rho(1900)$ and $\rho(2150)$, respectively, are input parameters
which are taken from PDG (see Table \ref{chidu}).

\begin{figure}[hbtp]
\centering
\includegraphics[width=0.48\textwidth,scale=0.2]{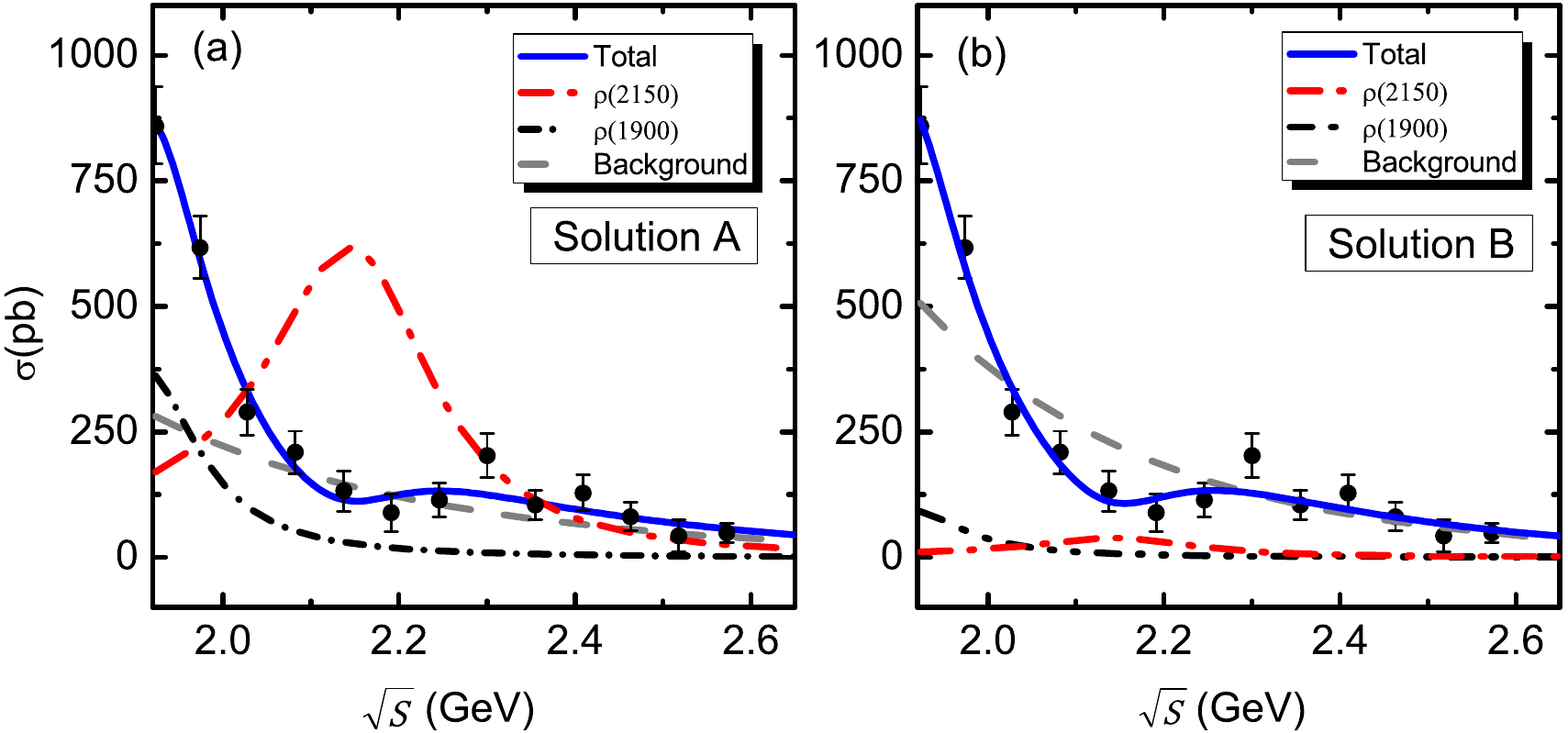}
\caption{(Color online.) The fitted result of the cross section of $e^+e^-\to\pi^+\pi^-$. Here, the black dots with error bar is {\it BABAR} result \cite{BABAR:2019oes}.
We present two solutions (Solution A (left) and Solution B (right)), which can depict the data well.}\label{scenario123}
\end{figure}

In the fitting process, we found that $\rho(2150)$ play dominant role to reproduce the line shape of $e^+e^-\to \pi^+\pi^-$ around 2.2 GeV \cite{BABAR:2019oes}.
We can find two solutions (solution A and solution B), both of which can reproduce the $BABAR$'s data well.
In Table \ref{para}, we list these obtained fitting parameters. And then, in Fig. \ref{scenario123}, we further present the fitted results and the comparison with the experimental data.
It is worth mentioning that the intermediate resonance $\rho(1900)$ has obvious contribution if describing the line shape corresponding to the center-of-mass energy $\sqrt{s}<2.2$ GeV.

We also notice obvious difference of the fitted $\Gamma_{e^+e^-}\mathcal{B}(\pi^+\pi^-)_{\rho(1900)}$ and $\Gamma_{e^+e^-}\mathcal{B}(\pi^+\pi^-)_{\rho(2150)}$ under two solutions, which makes us to
check the reasonability of the obtained $\Gamma_{e^+e^-}\mathcal{B}(\pi^+\pi^-)_{\rho(1900)}$ and $\Gamma_{e^+e^-}\mathcal{B}(\pi^+\pi^-)_{\rho(2150)}$ values.

$\Gamma_{e^+e^-}\mathcal{B}(\pi^+\pi^-)_{\rho(1900)}$ and $\Gamma_{e^+e^-}\mathcal{B}(\pi^+\pi^-)_{\rho(2150)}$ can be  theoretically calculated. In Ref. \cite{He:2013ttg}, Lanzhou group have
calculated the widths of these discussed $\rho$ mesons decaying into the $\pi^+\pi^-$ channel by the quark pair creation (QPC) model.
Under the framework of the potential model, the general expression of dilepton decay width of the discussed $\rho$ state \cite{Godfrey:1985xj} is
\begin{eqnarray}\label{eepipit}
\Gamma_{e^+e^-}=\frac{4\pi}{3}\alpha^2m_{\rho_i}\mathcal{M}_{\rho_i}^2,
\end{eqnarray}
where $\mathcal{M}_{\rho_i}$ denotes decay amplitude, which is defined as $\mathcal{M}_{\rho_i}=\sqrt{2}V_{\rho_i}$ and $\mathcal{M}_{\rho_i}=({4}/{3})^{1/2}V^{\prime}_{\rho_i}$
for $S-$wave and $D-$wave $\rho$ mesons, respectively.  Here, factor $V_{\rho_i}$ and $V^{\prime}_{\rho_i}$ read as
\begin{eqnarray}
\begin{split}
V_{\rho_i}=&m_{\rho_i}^{-2}\;\widetilde{m}_{\rho_i}^{1/2}\;(2\pi)^{3/2}\int d^3p\;(4\pi)^{-1/2}\phi_{\rho_i}(p)\;\left(\frac{m_1m_2}{E_1E_2}
\right)^{1/2},\\
V^{\prime}_{\rho_i}=&m_{\rho_i}^{-2}\;\widetilde{m}_{\rho_i}^{1/2}\;(2\pi)^{3/2}\int d^3p\;(4\pi)^{-1/2}\phi_{\rho_i}(p)\;\\
& \times\left(\frac{m_1m_2}{E_1E_2}
\right)^{1/2} \left(\frac{p}{E_1}\right)^2.
\end{split}
\end{eqnarray}
Here, $m_1=m_2=0.22$ GeV \cite{Godfrey:1985xj} is quark mass inside the $\rho$ meson, and $E_1$ and $E_2$  are the energy of the corresponding quarks. And then, $\widetilde{m}_{\rho_i}=2\int d^3p\;E\;|\phi_{\rho_i}(p)|^2$. The radial part of the spatial wave functions of these involved $\rho$ meson states can be
depicted by the radial part of simple harmonic oscillator (SHO) wave function  $\phi_{\rho_i}(p)$, i.e.,
\begin{eqnarray}\label{L2}
\begin{split}
\phi_{\rho_i}(p)=&(-1)^n(-i)^LR^{3/2}\;e^{-\frac{p^2R^2}{2}}\sqrt{\frac{2n!}{\Gamma(n+L+3/2)}}\left(pR\right)^L\\
&\times L_n^{L+1/2}\left(p^2R^2\right).
\end{split}
\end{eqnarray}
Here, $R$ that refers to the size of meson state is the parameter of SHO wave function and its possible value has been suggested by Ref. \cite{He:2013ttg}, which is summarized in Table \ref{chidu}.
Combined with the $\Gamma_{\pi^+\pi^-}$ listed in Table \ref{chidu}, $\Gamma_{e^+e^-}\mathcal{B}(\pi^+\pi^-)$ can be directly calculated, which is also dependent on $R$ value (see the seventh column in Table \ref{chidu}).
Since the fitted results of $\Gamma_{e^+e^-}\mathcal{B}(\pi^+\pi^-)_{\rho(1900)}$ and $\Gamma_{e^+e^-}\mathcal{B}(\pi^+\pi^-)_{\rho(2150)}$ are comparable with the calculated results from the potential model, the fitted result corresponding to Solution A in Fig. \ref{scenario123} and Table \ref{para} is more favorable.

\section{Summary}\label{sec5}

The $e^+e^-\to\pi^+\pi^-$ process is a good platform to study $\rho$-like states due to $G$-parity conservation.
Inspired by the {\it BABAR} measurement of the cross section of the $e^+e^-\to\pi^+\pi^-$ process around 2 GeV,
we study the contribution of higher radial excitations in the $\rho$ meson family to $e^+e^-\to\pi^+\pi^-$.  When reproducing the experimental data of $e^+e^-\to\pi^+\pi^-$ around 2.2 GeV,
$\rho(2150)$ and $\rho(1900)$ as $\rho(4S)$ and $\rho(3S)$ play important role. Combining with former study of mass spectrum and decay behavior of $\rho$ meson family \cite{He:2013ttg}, the present work enforces the assignment of $\rho(2150)$ and $\rho(1900)$ as $\rho(4S)$ and $\rho(3S)$, respectively, which is a crucial step in constructing $\rho$ meson family.

In recent years, BESIII measured some processes of $e^+e^-$ annihilation into light mesons \cite{Ablikim:2015orh,Ablikim:2018iyx,Ablikim:2019tpp,Ablikim:2020coo,Ablikim:2020pgw}. In the following, we may focus on other typical  processes of $e^+e^-$ annihilation into light mesons, which have close relation to light vector mesonic states. Since there exist abundant $\rho$, $\omega$ and $\phi$ higher excitations around 2 GeV,
studying these processes of $e^+e^-$ annihilation into light mesons is helpful to better understand the contribution of these light vector meson to these processes. Obviously, the present work is a beneficial attempt on this issue.

We also feel that promoting experimental precision can inspire theoretical progress. The present work is a good example for this point.
In the near future, BESIII and Belle II will be main force of the study of light hadron spectroscopy \cite{Ablikim:2019hff,Kou:2018nap}.  We also expect more precise experimental data of processes of $e^+e^-$ annihilation into light mesons, which is valuable to construct light vector meson family.

\section*{ACKNOWLEDGEMENTS}

This work is supported by the China National Funds for Distinguished Young Scientists under Grant No. 11825503, the National Program for Support of Top-notch Young Professionals and and the projects funded by Science and Tech- nology Department of Qinghai Province No. 2020-ZJ-728.

\vfil



\begin{thebibliography}{90}


\bibitem{Augustin:1974xw}
J.~Augustin \textit{et al.} [SLAC-SP-017],
Discovery of a Narrow Resonance in $e^+ e^-$ Annihilation,
Phys. Rev. Lett. \textbf{33}, 1406-1408 (1974).

\bibitem{Aubert:2005rm}
B.~Aubert \textit{et al.} [BaBar],
Observation of a broad structure in the $\pi^+ \pi^- J/\psi$ mass spectrum around 4.26-GeV/$c^2$,
Phys. Rev. Lett. \textbf{95}, 142001 (2005).

\bibitem{Lees:2012cj}
J.~Lees \textit{et al.} [BaBar],
Precise Measurement of the $e^+ e^- \to \pi^+\pi^- (\gamma)$ Cross Section with the Initial-State Radiation Method at {\it BABAR},
Phys. Rev. D \textbf{86}, 032013 (2012).


\bibitem{BABAR:2019oes}
J.~Lees \textit{et al.} [BaBar],
Resonances in $e^+e^-$ annihilation near 2.2 GeV,
Phys. Rev. D \textbf{101}, no.1, 012011 (2020).

\bibitem{Wang:2012wa}
X.~Wang, Z.~Sun, D.~Chen, X.~Liu and T.~Matsuki,
Non-strange partner of strangeonium-like state $Y(2175)$,
Phys. Rev. D \textbf{85}, 074024 (2012).

\bibitem{He:2013ttg}
L.~He, X.~Wang and X.~Liu,
Towards two-body strong decay behavior of higher $\rho$ and $\rho_3$ mesons,
Phys. Rev. D \textbf{88}, no.3, 034008 (2013).


\bibitem{Wang:2019jch}
C.~Pang, Y.~Wang, J.~Hu, T.~Zhang and X.~Liu,
Study of the $\omega$ meson family and newly observed $\omega$-like state $X(2240)$,
Phys. Rev. D \textbf{101}, 074022 (2020).

\bibitem{Tanabashi:2018oca}
  M.~Tanabashi {\it et al.} [Particle Data Group],
  Review of Particle Physics,
  Phys.\ Rev.\ D {\bf 98}, no. 3, 030001 (2018).

\bibitem{Godfrey:1985xj}
  S.~Godfrey and N.~Isgur,
  Mesons in a Relativized Quark Model with Chromodynamics,
  Phys.\ Rev.\ D {\bf 32}, 189 (1985).

\bibitem{Abele:2001pv}
  A.~Abele {\it et al.} [CRYSTAL BARREL Collaboration],
  $4\pi$ decays of scalar and vector mesons,
  Eur.\ Phys.\ J.\ C {\bf 21}, 261 (2001).

\bibitem{Kittimanapun:2008wg}
  K.~Kittimanapun, Y.~Yan, K.~Khosonthongkee, C.~Kobdaj and P.~Suebka,
  $e^+e^-\to\omega\phi$ reaction and rho(1450) and rho(1700) mesons in a quark model,
  Phys.\ Rev.\ C {\bf 79}, 025201 (2009).


\bibitem{Castro:1988hp}
  A.~Castro {\it et al.} [DM2 Collaboration],
  The $\pi$, $K$, Proton Electromagnetic Form-factors And New Related Dm2 Results,
  LAL-88-58.

\bibitem{Antonelli:1996xn}
  A.~Antonelli {\it et al.} [FENICE Collaboration],
  Measurement of the total $e^+e^-\to\mathrm{hadrons}$ cross-section near the $e^+e^-\to N\bar{N}$ threshold,
  Phys.\ Lett.\ B {\bf 365}, 427 (1996).

\bibitem{Frabetti:2001ah}
  P.~L.~Frabetti {\it et al.} [E687 Collaboration],
  Evidence for a Narrow Dip Structure at 1.9-GeV/$c^{2}$ in 3$\pi^{+} 3\pi^{-}$ Diffractive photoproduction,
  Phys.\ Lett.\ B {\bf 514}, 240 (2001).

\bibitem{Frabetti:2003pw}
  P.~L.~Frabetti {\it et al.},
  On the narrow dip structure at 1.9-GeV$/c^2$ in diffractive photoproduction,
  Phys.\ Lett.\ B {\bf 578}, 290 (2004).


\bibitem{Aubert:2006jq}
  B.~Aubert {\it et al.} [BaBar Collaboration],
  The $e^+e^- \to 3(\pi^+ \pi^-), 2(\pi^+ \pi^- \pi^0)$ and $K^+ K^- 2(\pi^+ \pi^-)$ cross sections at center-of-mass energies from production threshold to 4.5GeV measured with initial-state radiation,
  Phys.\ Rev.\ D {\bf 73}, 052003 (2006).


\bibitem{Aubert:2007ym}
  B.~Aubert {\it et al.} [BaBar Collaboration],
  Measurements of $e^{+} e^{-} \to K^{+} K^{-} \eta$, $K^{+} K^{-} \pi^0$ and $K^0_{s} K^\pm \pi^\mp$ cross- sections using initial state radiation events,
  Phys.\ Rev.\ D {\bf 77}, 092002 (2008).

\bibitem{Solodov:2011dn}
  E.~P.~Solodov [CMD-3 Collaboration],
  First results from the CMD3 Detector at the VEPP2000 Collider,
  arXiv:1108.6174 [hep-ex].

\bibitem{Surovtsev:2008zza}
  Y.~S.~Surovtsev and P.~Bydzovsky,
  Analysis of the pion pion scattering data and rho-like mesons,
  Nucl.\ Phys.\ A {\bf 807}, 145 (2008).
  
\bibitem{Bugg:2012yt}
  D.~V.~Bugg,
  Comment on "Systematics of radial and angular-momentum Regge trajectories of light nonstrange $q\overline{q}$-states",
  Phys.\ Rev.\ D {\bf 87}, no. 11, 118501 (2013).

\bibitem{Barnes:1996ff}
  T.~Barnes, F.~E.~Close, P.~R.~Page and E.~S.~Swanson,
  Higher quarkonia,
  Phys.\ Rev.\ D {\bf 55}, 4157 (1997).

\bibitem{Bisello:1981sh}
  D.~Bisello, J.~C.~Bizot, J.~Buon, A.~Cordier, B.~Delcourt and F.~Mane,
  Study of the Reaction $e^+ e^- \to 3 \pi^+ 3 \pi^-$ in the Total Energy Range 1400-{MeV} to 2180-{MeV},
  Phys.\ Lett.\  {\bf 107B}, 145 (1981).

\bibitem{Atkinson:1985yx}
  M.~Atkinson {\it et al.} [Omega Photon Collaboration],
  Evidence for a $\omega \rho^\pm \pi^\mp$ State in Diffractive Photoproduction,
  Z.\ Phys.\ C {\bf 29}, 333 (1985).

\bibitem{Clegg:1989mp}
  A.~B.~Clegg and A.~Donnachie,
  $\rho^\prime$s in 6 $\pi$ States From Materialization of Photons,
  Z.\ Phys.\ C {\bf 45}, 677 (1990).

\bibitem{Biagini:1990ze}
  M.~E.~Biagini, S.~Dubnicka, E.~Etim and P.~Kolar,
  Phenomenological evidence for a third radial excitation of $\rho$(770),
  Nuovo Cim.\ A {\bf 104}, 363 (1991).

\bibitem{Alde:1992wv}
  A.~Alde {\it et al.} [IHEP-IISN-LANL-LAPP-KEK Collaboration],
  Study of the $\omega\pi_0$ system,
  Z.\ Phys.\ C {\bf 54}, 553 (1992).

\bibitem{Alde:1994jm}
  D.~Alde {\it et al.} [GAMS Collaboration],
  Partial wave analysis of the $\omega\pi_0$ system at high masses,
  Nuovo Cim.\ A {\bf 107}, 1867 (1994)
  [Z.\ Phys.\ C {\bf 66}, 379 (1995)].

\bibitem{Anisovich:2000ut}
  A.~V.~Anisovich {\it et al.},
 $ I = 0$ $C = +1$ mesons from 1920 to 2410 MeV,
  Phys.\ Lett.\ B {\bf 491}, 47 (2000).

\bibitem{Anisovich:2001vt}
  A.~V.~Anisovich, C.~A.~Baker, C.~J.~Batty, D.~V.~Bugg, V.~A.~Nikonov, A.~V.~Sarantsev, V.~V.~Sarantsev and B.~S.~Zou,
  Resonances in $p\bar{p}\to\omega\eta\pi_0$ in the mass range 1960-MeV to 2410-MeV,
  Phys.\ Lett.\ B {\bf 513}, 281 (2001).

\bibitem{Anisovich:1999xm}
  A.~V.~Anisovich {\it et al.},
  Analysis of $p\bar{p}\to\pi^+\pi^-$, $\pi^0\pi^0$, $\eta\eta$ and $\eta\eta'$ from threshold to 2.5 GeV/$c$,
  Phys.\ Lett.\ B {\bf 471}, 271 (1999).

\bibitem{Anisovich:2002su}
  A.~V.~Anisovich {\it et al.},
  Combined analysis of meson channels with $I = 1$, $C = -1$ from 1940 to 2410 MeV,
  Phys.\ Lett.\ B {\bf 542}, 8 (2002).


\bibitem{Aubert:2007ef}
  B.~Aubert {\it et al.} [BaBar Collaboration],
  The $e^+e^-\to 2(\pi^+\pi^-)\pi^0$, $2(\pi^+\pi^-)\eta$, $K^+K^-\pi^+\pi^-\pi^0$ and $K^+K^-\pi^+\pi^-\eta$ Cross Sections Measured with Initial-State Radiation,
  Phys.\ Rev.\ D {\bf 76}, 092005 (2007)
  Erratum: [Phys.\ Rev.\ D {\bf 77}, 119902 (2008)].

\bibitem{Anisovich:2000kxa}
  A.~V.~Anisovich, V.~V.~Anisovich and A.~V.~Sarantsev,
  Systematics of $q\bar{q}$ states in the ($n$,$M^2$) and ($J$, $M^2$) planes,
  Phys.\ Rev.\ D {\bf 62}, 051502 (2000).

\bibitem{Masjuan:2012gc}
  P.~Masjuan, E.~Ruiz Arriola and W.~Broniowski,
  Systematics of radial and angular-momentum Regge trajectories of light non-strange $q\bar{q}$-states,
  Phys.\ Rev.\ D {\bf 85}, 094006 (2012).

\bibitem{Masjuan:2013xta}
  P.~Masjuan, E.~Ruiz Arriola and W.~Broniowski,
  Reply to "Comment on 'Systematics of radial and angular-momentum Regge trajectories of light nonstrange $q\overline{q}$-states'",
  Phys.\ Rev.\ D {\bf 87}, no. 11, 118502 (2013)


\bibitem{Hasan:1994he}
  A.~Hasan and D.~V.~Bugg,
  Amplitudes for $p\bar{p}\to\pi\pi$ from 0.36-GeV/c to 2.5-GeV/$c$,
  Phys.\ Lett.\ B {\bf 334}, 215 (1994).

\bibitem{Chen:2020xho}
  D.~Y.~Chen, J.~Liu and J.~He,
  Reconciling the $X$(2240) with the $Y$(2175),
  Phys.\ Rev.\ D {\bf 101}, no. 7, 074045 (2020).


\bibitem{Bugg:2003kj}
  D.~V.~Bugg,
  Comments on the $\sigma$ and $\kappa$,
  Phys.\ Lett.\ B {\bf 572} (2003) 1
   Erratum: [Phys.\ Lett.\ B {\bf 595} (2004) 556].

\bibitem{Ablikim:2015orh}
  M.~Ablikim {\it et al.} [BESIII Collaboration],
  Measurement of the $e^+ e^- \to \pi^+ \pi^-$ cross section between 600 and 900 MeV using initial state radiation,
  Phys.\ Lett.\ B {\bf 753}, 629 (2016).

\bibitem{Ablikim:2018iyx}
  M.~Ablikim {\it et al.} [BESIII Collaboration],
  Measurement of $e^{+} e^{-} \rightarrow K^{+} K^{-}$ cross section at $\sqrt{s} = 2.00 - 3.08$ GeV,
  Phys.\ Rev.\ D {\bf 99}, no. 3, 032001 (2019).

\bibitem{Ablikim:2019tpp}
  M.~Ablikim {\it et al.} [BESIII Collaboration],
  Cross section measurements of $e^{+}e^{-} \to K^{+}K^{-}K^{+}K^{-} $ and $ \phi K^{+}K^{-}$ at center-of-mass energies from 2.10 to 3.08 GeV,
  Phys.\ Rev.\ D {\bf 100}, no. 3, 032009 (2019).

\bibitem{Ablikim:2020coo}
  M.~Ablikim {\it et al.} [BESIII Collaboration],
  Observation of a structure in $e^{+}e^{-} \to \phi \eta^{\prime}$ at $\sqrt{s}$ from 2.05 to 3.08 GeV,
  arXiv:2003.13064 [hep-ex].

\bibitem{Ablikim:2020pgw}
M.~Ablikim \textit{et al.} [BESIII],
Observation of a Resonant Structure in $e^{+}e^{-} \to K^{+}K^{-}\pi^{0}\pi^{0}$,
Phys. Rev. Lett. \textbf{124}, no.11, 112001 (2020).

\bibitem{Ablikim:2019hff}
  M.~Ablikim {\it et al.},
  Future Physics Programme of BESIII,
  Chin.\ Phys.\ C {\bf 44}, no. 4, 040001 (2020).

\bibitem{Kou:2018nap}
  E.~Kou {\it et al.} [Belle-II Collaboration],
  The Belle II Physics Book,
  PTEP {\bf 2019}, no. 12, 123C01 (2019)
  Erratum: [PTEP {\bf 2020}, no. 2, 029201 (2020)].

\end{thebibliography}
\end{document}